# Solar Power Generation Profile Estimation for Lunar Surface Solar PV Systems

Jesus Silva-Rodriguez and Xingpeng Li

*Abstract*—As NASA prepares to carry out its Artemis lunar missions, the design and planning of robust power systems tailored to the lunar environment become necessary and urgent. Solar photovoltaic (PV) systems are among the most suitable power generators for lunar applications given the abundant solar irradiance the lunar surface receives as a result of the lack of an atmosphere. However, the vastly different environmental conditions of the moon compared to those on Earth call for special reconsiderations to the traditional PV power output models used for terrestrial applications. The substantially wider range of temperatures on the moon combined with the absence of atmosphere implies the PV will operate at very different thermal conditions. Therefore, this paper proposes a PV power output model that determines PV cell temperature on the lunar surface based on lunar ambient temperature as well as solar irradiance, while also capturing these special lunar conditions. The model is based on thermal exchange via the three main heat transfer paths: heat radiation, absorption, and convection. Moreover, four different array configurations are presented and compared to determine the most suitable choice based on energy generation and complexity for different locations around the moon.

*Index Terms*--Lunar Surface, Profile Estimation, PV Systems, PV Thermal Model, Solar Power.

## I. INTRODUCTION

As NASA's Artemis program continues to progress forward, NASA is preparing to launch a second lunar landing by 2024, followed by additional launch missions to carry out continuous and sustainable lunar exploration and habitation by the mid to late 2020s [1]. For these Artemis missions and future plans to be possible, especially when it comes to human settlements on the moon, a robust power system with its own sustainable power sources is necessary.

A solar photovoltaic (PV) system is a type of power generator that can offer the most abundant energy source on the Moon due to the negligible atmosphere present which in turn causes essentially unobstructed and non-diffused reception of solar radiation on the lunar surface [2], [3], as opposed to the solar radiation received by the Earth surface which is attenuated by our dense terrestrial atmosphere. Therefore, solar power generation systems such as PV arrays can be the most advantageous power generators for lunar facilities.

Solar PV systems for power generation on lunar soil have already been considered and theoretically analyzed. NASA has been studying these technologies for a few decades and has already proposed solar PV cells and energy storage technologies as candidates to power lunar bases, as well as some fixed array orientations and configurations [4], in addition to other potential generation technologies such as fission nuclear reactors like the Kilopower project [5]. However, so far, any solar tracking technology for PV arrays has been left out to keep systems and components as simple as possible to facilitate payload shipments [4].

Moreover, the Artemis Base Camp is planned to be deployed at the south pole of the Moon [1], where there is substantial supporting evidence that some south polar locations exhibit prolonged time periods of constant sunlight illumination [3], [6]. Therefore, solar PV systems would be even more beneficial in south polar regions, and would help support reliable power generation. Still, due to the Moon's angle of declination and rotation, lunar seasonal periods exist much like those we experience on Earth, changing the duration of daytime and nighttime periods [6], [7]. This means that even at the south pole, there exists time periods in which some polar locations will experience zero illumination (i.e., nighttime). Therefore, in order to devise a reliable power consumption schedule for the different activities of the Artemis Base Camp, establishing adequate estimates of power generation profiles is of crucial importance. This can ensure that the most critical power loads can be supplied even at times of low power generation, and all activities can be planned accordingly.

Some recent studies have already developed potential ways to compute estimates of solar illumination profiles at different lunar south polar locations with different terrain elevations [3], [8]. In [3], an algorithm is developed to seek the most optimal configuration for PV systems at different sites, placed on top of towers of different altitudes and interconnected via power transmission lines of different distances in an attempt to leverage the illumination differences throughout the lunar seasons at different locations of the south pole. The algorithm determines solar illumination based on computed Sun angles of elevation on the lunar sky as well as lunar topographic information. This technique to determine solar illumination profiles is reutilized in [8] with the objective of determining solar power generation profiles based on PV systems on the moon, using traditional PV power generation theory that relates power output to solar irradiance intensity, PV panel orientation, PV efficiency, inclination angles, and PV surface area. However, none of these papers consider solar power generation as a function of temperature, when it is well known that there exists a temperature dependence of PV system efficiency [4], [9]. Additionally, substantial ambient temperature variations during lunar diurnal cycles can also affect PV cell temperature, and in turn their operations and performance [5], [10].

Therefore, this paper proposes a PV power output model for lunar surface applications as a function of collected solar

---

Jesus Silva-Rodriguez and Xingpeng Li are with the Department of Electrical and Computer Engineering, University of Houston, Houston, TX, 77204, USA (e-mail: jasilvarodriguez@uh.edu; xingpeng.li@asu.edu).

irradiance and ambient temperature of the lunar surroundings, with the purpose of developing solar power generation profiles based on local latitude. Additionally, this paper also presents a comparison of four different array orientation configurations, including axial rotational tracking, to determine the total energy and power output each configuration can achieve at different latitudes.

The rest of the paper is organized as follows. Section II covers different lunar environmental aspects relevant to solar power generation. Section III presents the mathematical theory used to determine incident irradiance angles on solar PV panels for the different array configurations studied in this paper, as well as the thermal model used to determine PV cell temperature as a function of ambient temperature and collector irradiance. Section IV compares the net energy generations and presents solar power output profiles for each array configuration at different lunar latitudes. Lastly, Section V concludes the paper.

## II. LUNAR ENVIRONMENT

The power generation profiles for solar PV systems on the Moon will depend on levels of solar irradiance and surrounding ambient temperature at those locations. Additionally, because of special conditions on the lunar surface, it appears the lunar ground temperature is directly proportional to the solar radiation striking the ground [10]. Therefore, determining ground irradiance profiles is crucial to estimate ambient temperature profiles, and in turn PV power generation profiles.

Moreover, the level of intensity of solar irradiance depends on the sun's position in the sky, which in turn depends on the local time of a lunar day, as well as the location [11]. For instance, at sunrise and sunset, the sun's position in the sky is exactly at the horizon. Therefore, to be able to determine levels of solar irradiance intensity, knowing where the sun is in the sky at any time of the lunar day is of great importance as well.

Another important parameter that influences the sun's position in the sky is the axis declination angle. For the Moon's case, the rotational axis has a maximum declination angle of 1.54° [6], and it varies in a sinusoidal pattern with a period of about 346.7 earth-days [7], which corresponds to a lunar year. Fitting the data of [7] to a sinusoidal pattern creates the periodic sine function represented by (1), which is illustrated in Fig. 1.

$$\delta = \tau \sin\left[\frac{360°}{Y}\left(n + \frac{t}{24} - t_0\right)\right], \quad (1)$$

where $\tau$ = -1.545°, $Y$ = 346.71 earth-days, $t_0$ = -1.23 earth-days.

To adjust (1) to the data, the term $n$ is defined to be a day number, with $n = 1$ representing the date of the beginning of the data, that is Jan 1st, 2020, in this case, and $t$ is the time variable representing the time of the lunar day in hours. In this way, the lunar declination angle $\delta$ can be calculated for any date in the future and be used to estimate the sun's position in the lunar sky.

### A. Solar Elevation Angle

The solar elevation angle is the angle between the sun and the local horizon directly beneath the sun's position in the sky. It depends on the local latitude, time of day, and declination angle. Mathematically, it is defined as follows [11]:

$$\beta = \sin^{-1}[\cos L \cos \delta \cos H + \sin L \sin \delta], \quad (2)$$

where $L$ represents the local latitude on the lunar surface, and $H$ is the "hour angle". The hour angle is defined as the number of degrees the moon must rotate before the sun is directly over the local meridian, which represents the local noon. Considering that a lunar day lasts around 708.75 hours [6], [8], then it can be stated that the moon rotates at an angular speed of 0.515° per hour. Therefore, the lunar hour angle can be defined as follows:

$$H = 0.515\frac{°}{hr} \cdot (354.365 \ hrs - t), 0 \leq t < 708.75 \ hrs. \quad (3)$$

The hour being substituted into variable $t$ must then be adjusted to coincide with the lunar diurnal cycle, with $t = 0$ representing the local noon. This adjustment must also be made for (1).

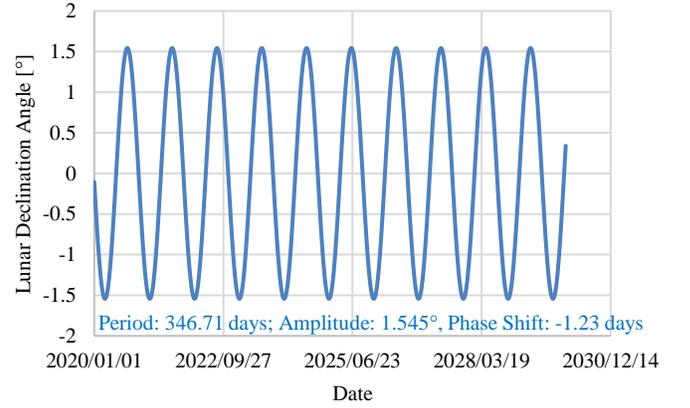

Fig. 1: Lunar declination angle for different date ranges.

### B. Solar Azimuth Angle

The solar azimuth angle is the angle between the sun's position in the sky projected into the horizontal and either the north or the south direction. Similar to the solar elevation angle, it also depends on local latitude, time of day, and the declination angle. It is defined as follows [11]:

$$\varphi_S = \sin^{-1}\left[\frac{\cos \delta \sin H}{\cos \beta}\right]. \quad (4)$$

By convention, the solar azimuth angle is defined to be positive before local noon with the sun in the east and negative after local noon with the sun in the west. It is measured relative to due south in the northern hemisphere and relative to due north in the southern hemisphere [11]. Fig. 2 shows a diagram depicting both solar elevation and azimuth angles.

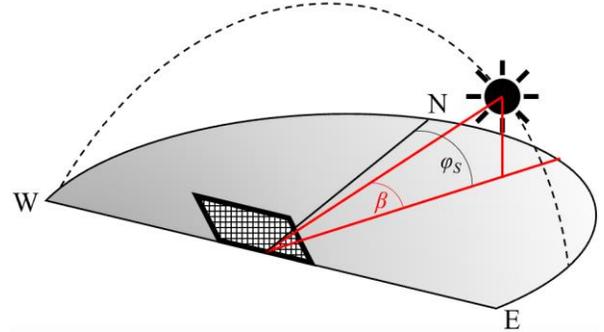

Fig. 2: Solar elevation and solar azimuth angles based on sun's position in the sky.

### C. Horizontal Irradiance

For the scope of this paper, all locations considered for the placement of solar PV systems are assumed to be completely horizontal, perpendicular to the diameter of the moon. Therefore, the horizontal irradiance is determined assuming a surface that is tilted at 0°. The angle of incidence of the solar irradiance is a function of multiple factors pertaining to the position and orientation of the surface being considered. In general, this angle of incidence is mathematically determined with the following expression [11]:

$$\theta = \cos^{-1}[\cos\beta \cos(\varphi_S - \varphi_C)\sin\Sigma + \sin\beta\cos\Sigma], \quad (5)$$

where $\Sigma$ represents the tilt angle of the surface and $\varphi_C$ the azimuth of the surface horizontal orientation.

Substituting the tilt angle $\Sigma$ with 0° for the special case of a horizontal surface, (5) can be simplified as follows:

$$\theta = \cos^{-1}[\sin\beta]. \quad (6)$$

Therefore, the angle of incidence for a horizontal surface will exclusively depend on the solar elevation angle.

The net irradiance intensity a horizontal surface receives is then dependent on the angle of incidence θ [11], and for this case it can be calculated as follows:

$$I_{BH} = I_B \cos\theta = I_B \sin\beta, \quad (7)$$

where $I_B$ represents the total solar radiation received by the moon's surface, which is found to be the same as the radiation intensity received just outside Earth's atmosphere: 1361 W/m² [12]. This is because of the absence of an atmosphere on the moon that could attenuate this radiation intensity such as on Earth [3].

Using (1)-(7), horizontal irradiance profiles can be derived for different locations at different dates and times of the lunar day. As an example, Fig. 3 shows what the horizontal irradiance would be at latitudes near the center of the south pole for a single lunar day (~708 hours). The profile shows how the horizontal irradiance starts to behave on a sinusoidal pattern of decreasing minimum points over time, but with a larger magnitude variation as the location moves away from the center of the south pole. This behavior occurs as a result of the maximum lunar axis tilt of 1.54°, and the polar latitudes having a more indirect reception of the solar rays.

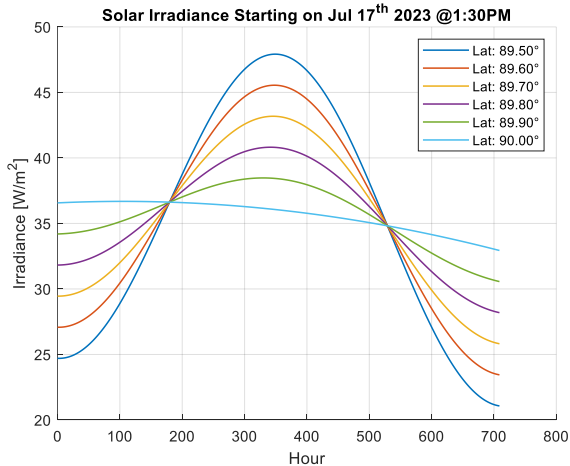

Fig. 3: Horizontal solar irradiance profiles for a lunar day at different south polar latitudes.

### D. Ground Temperature

Change in temperature of materials is directly proportional to the energy they absorb. Granted that realistically not all of the energy absorbed by the lunar surface will come from solar radiation, its temperature will still be mainly proportional to it, especially for the lack of atmosphere that prevents any heat exchange with the surface and any possible surrounding gases or winds [2]. This direct relation between solar irradiance and surface temperature is proven in [5] and [10], where temperature statistics presented show that even in adjacent areas near south polar craters the temperature varies drastically between illuminated areas and shadowed areas, with temperature differences of around 300 K between both areas.

Surface temperature profiles can then be estimated by assuming a direct relation of temperature with incoming solar irradiance levels. In this paper, temperature statistics for locations around the Shackleton crater are taken from [10], and they are used to map the maximum and minimum temperature points to the horizontal irradiance profiles.

The following expression is used to relate lunar ambient temperature to horizontal irradiance:

$$T_a = (T_{a,max} - T_{a,min}) \cdot \left(\frac{I_{BH} - I_{BH,min}}{I_{BH,max} - I_{BH,min}}\right) + T_{a,min}, \quad (8)$$

where $I_{BH,min}$ and $I_{BH,max}$ are the minimum and maximum horizontal irradiance magnitudes of the profile respectively, and $T_{a,min}$ and $T_{a,max}$ are the minimum and maximum ground temperatures of the area of interest respectively.

## III. LUNAR PV SYSTEMS

Once the surrounding ambient irradiance and temperature are known, the last portion to determine solar power generation would be the position and orientation of the solar PV collectors.

As it was mathematically described in (5), the angle of incidence on any surface depends on not only the sun's position in the sky but also the horizontal and vertical angular positions of the surface's orientation. Moreover, for a non-horizontal surface (a surface with a tilt angle $\Sigma \neq 0$), the total irradiance received will be the combination of the direct irradiance from the sun and any irradiance being reflected by the surroundings' surfaces, which in this case correspond to the lunar ground. Mathematically, the PV collected irradiance is given as follows:

$$I_C = I_{BC} + I_{RC} \quad (9)$$

with,

$$I_{BC} = I_B \cos\theta \quad (10)$$

$$I_{RC} = \rho I_{BH}\left(\frac{1-\cos\Sigma}{2}\right). \quad (11)$$

where the ρ term corresponds to the ground reflectance, which for the scope of this paper is assumed to be constant at a value of 0.2, per the data presented in [13].

The solar elevation and azimuth angles are based on their respective profiles during a lunar year, and can be used as parameters to determine optimal array orientation configurations. The variables to be determined will be the PV collector tilt angle and horizontal orientation azimuth, also known as collector azimuth, when working with PV systems. This paper compares four different configurations which are presented and described in detail in the following subsection.

### A. Array Orientation Configurations

#### 1) Fixed Array

The fixed array configuration assumes the orientation of both collector azimuth and tilt angles are fixed, with the collector azimuth always facing the equator and the tilt angle set to be equal to the latitude of the location. This configuration corresponds to the optimal fixed angular position values to maximize solar irradiance exposure [11]. Thus, this configuration simply assumes $\phi_C = 0°$ and $\Sigma = L$.

*2) Triangular Array*

This configuration places the PV panels of an array in a "tent-like" form, with half the panels facing the east ($\phi_C = 90°$) and the other half facing the west ($\phi_C = -90°$), tilted at an angle $\Sigma = \alpha$ from the horizontal [4], as roughly depicted in Fig. 4.

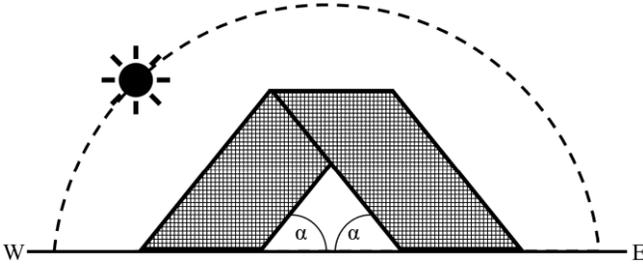

Fig. 4: Triangular array configuration diagram.

For this configuration, each half of the array will experience its own angle of incidence $\theta$ that contributes to the net solar irradiance collected by the array, which are expressed in (12) and (13) respectively.

$$\theta_W = \cos^{-1}[\cos\beta \cos(\varphi_S + 90°)\sin\alpha + \sin\beta \cos\alpha] \quad (12)$$

$$\theta_E = \cos^{-1}[\cos\beta \cos(\varphi_S - 90°)\sin\alpha + \sin\beta \cos\alpha]. \quad (13)$$

The direct irradiance is then calculated with these angles of incidence combined in the following way:

$$I_{BC} = I_B \cdot [\cos\theta_W + \cos\theta_E]. \quad (14)$$

According to [4], an "equilateral tent array," meaning an angle $\alpha = 60°$, achieves the most uniform power generation profile. Therefore, this angle is adopted as the tilt angle for this configuration in this paper.

*3) Horizontal Tracking with Fixed Tilt*

Horizontal tracking means that the collector azimuth angle $\phi_C$ can be manipulated to seek optimization of the angle of incidence at every time interval. Inspecting (5), it can be observed that to maximize $\theta$ controlling $\phi_C$, it must be set to equal the solar azimuth angle $\phi_S$ over all time periods, that is, to track the horizontal projection of the sun's position in the sky onto the ground. Assuming perfect horizontal tracking of the solar azimuth, (5) is simplified to the following expression for this array configuration:

$$\theta = \cos^{-1}[\cos\beta \sin\Sigma + \sin\beta \cos\Sigma], \quad (15)$$

which can be simplified further using trigonometric identities, taking the following simpler form:

$$\theta = \cos^{-1}[\sin(\beta + \Sigma)]. \quad (16)$$

This expression will always work as long as $\phi_C = \phi_S$ at all times.

Now it is only a matter of finding the optimal tilt angle $\Sigma$ that can maximize collected irradiance for an entire lunar year. To accomplish this, it is necessary to reconsider the expression for direct solar irradiance. With perfect horizontal tracking, (10) can take the following form for every time interval:

$$I_{BC,t} = I_B \sin(\beta_t + \Sigma). \quad (17)$$

However, simply maximizing the sum of $I_{BC}$ for all $t$ in the lunar year could provide erroneous values at time intervals when the solar altitude $\beta$ is negative, indicating nighttime. At these times, the irradiance must be zero since the sun is below the horizon, unable to illuminate the PV panels. Therefore, (17) must be conditioned, and redefined in the following way:

$$I_{BC,t} = \begin{cases} I_B \sin(\beta_t + \Sigma), & \beta_t \geq 0 \\ 0, & \beta_t < 0 \end{cases} \quad (18)$$

Maximizing the sum of $I_{BC}$ over the lunar year seems to indicate that optimal fixed $\Sigma$ values depend on latitude. Therefore, to enable perfect horizontal tracking of the solar azimuth, the fixed tilt angle must be selected according to Fig. 5, which was obtained by maximizing energy generation for the entire lunar year at every lunar latitude.

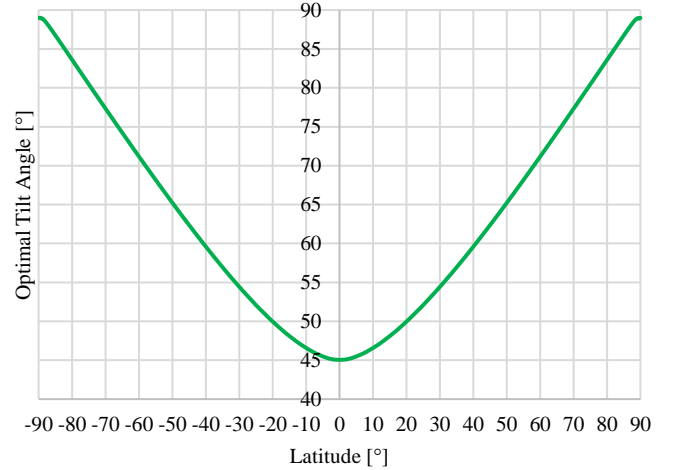

Fig. 5: Optimal fixed tilt angle values at different latitudes for horizontal tracking with fixed tilt array configuration.

*4) Perfect Horizontal and Vertical Tracking*

This configuration follows the same theory as before to enable horizontal tracking of the solar azimuth angle. However, instead of fixing the tilt of the arrays to an optimal value it allows it to be controlled as well to further optimize the angle of incidence, and hence the electrical energy generation. Inspection of (16) reveals that if $\theta$ wishes to be optimized further, then the tilt $\Sigma$ simply needs to always complement the solar altitude angle $\beta$ at all times, meaning that,

$$\Sigma = 90° - \beta \quad (19)$$

In conclusion, by setting the collector azimuth angle equal to the solar azimuth angle ($\phi_C = \phi_S$) and the tilt angle equal to the complement of the solar altitude angle such as in (19), then perfect horizontal and vertical tracking of the sun's position in the sky is achieved.

*B. PV Cell Temperature*

As it is well known, solar PV cell performance decreases with increasing temperatures [4], indicating a clear dependence of PV efficiency with cell temperature, which in fact has been well studied in literature [9], [14], [15]. PV efficiency can then be estimated with a linear regression based on rated reference values as follows:

$$\eta_{pv} = \eta_{ref} \cdot [1 - \gamma_{ref}(T_{cell} - T_{ref})] \quad (19)$$

where $\eta_{\text{ref}}$, $\gamma_{\text{ref}}$, and $T_{ref}$ are the rated efficiency, temperature relation coefficient, and rated cell temperature of the PV module, respectively. This information is normally provided by PV manufacturers for their modules.

For Earth applications, the cell temperature $T_{cell}$ can be determined using already established models based on standard testing conditions (STC) and nominal operating conditions for numerous different ambient scenarios on Earth [14]-[16]. However, the same assumptions are not compatible for lunar applications given the drastic environmental differences such as the absence of an atmosphere, and hence absence of wind, and the drastically different temperatures and higher solar radiation levels. As a result, it is necessary to derive a PV cell temperature model tailored for lunar conditions which can relate cell temperature to the temperature of the surroundings.

To construct a more accurate model for PV cell temperature on the moon, a thermal model that accounts for all present paths of heat transfer needs to be derived. To accomplish this, a heat balance equation can be used, which for PV modules will involve three main heat transfer paths: heat radiation $q_r$, heat absorption $q_s$, and heat convection $q_c$. These are related to PV power output as follows [17], [18]:

$$C_{pv}\frac{dT_{cell}}{dt} = q_s + q_c + q_r - P_{pv}, \quad (20)$$

where $C_{pv}$ is the PV module's heat capacity.

Considering the lunar environment, the convective heat rate can be taken as zero all the time ($q_c = 0$) due to the absence of an atmosphere since there is no air that can exchange heat with the PV cells [2]. Moreover, assuming a steady-state thermal model as well removes the differential term since only individual time instances are considered in this paper (15-min time intervals). These assumptions help simplify the heat balance equation, simplifying (20) to now be,

$$P_{pv} = q_s + q_r. \quad (21)$$

The heat absorption rate $q_s$ and heat radiation rate $q_r$ for solar PV cells are given as follows [18]:

$$q_s = \alpha_{cell} I_C A \quad (22)$$

$$q_r = A\sigma\left[\left(\frac{1+\cos\Sigma}{2}\right)\varepsilon_{sky}T_{sky}^4 + \left(\frac{1-\cos\Sigma}{2}\right)\varepsilon_a T_a^4 - \varepsilon_{cell}T_{cell}^4\right], \quad (23)$$

where $\alpha_{cell}$ is the PV cell absorptivity; $A$ is the PV cell surface area; $\sigma$ is the Stefan-Boltzmann constant (5.669 x 10⁻⁸ W/(m²K⁴)); $\varepsilon_{sky}$ is the emissivity of the lunar sky; $T_{sky}$ is the atmospheric temperature of the sky; $\varepsilon_a$ is the emissivity of the surroundings (i.e., the lunar ground); $T_a$ is the temperature of the surroundings (i.e., ambient temperature); and $\varepsilon_{cell}$ is the PV cell emissivity.

The absence of an atmosphere allows for the temperature of the sky to be taken as zero since the sky in this case can be considered to be open space directly. Therefore, with this and the previous lunar environmental-specific assumptions and using (21)-(23), the resulting function that relates cell temperature with ambient temperature and collector irradiance can be presented as follows:

$$P_{pv} = A\sigma\left[\left(\frac{1-\cos\Sigma}{2}\right)\varepsilon_a T_a^4 - \varepsilon_{cell}T_{cell}^4\right] + \alpha_{cell}I_C A \quad (24)$$

Now it is only needed to define PV power output $P_{pv}$ in terms of cell temperature as well to fully derive an expression that can estimate $T_{cell}$ as a bivariate function of $T_a$ and $I_C$.

*C. Lunar PV Power Output Model*

Solar PV cells convert incoming solar irradiance to power output in relation to their surface area and efficiency, as given by the following expression [9], [19]:

$$P_{pv} = \eta_{pv} A I_C. \quad (25)$$

However, solar PV systems are rarely described by their surface area size, and instead they are described by their rated power, normally determined experimentally under STC [9]. Based on rated power and other defined rated PV parameters, an expression representing a linear proportion between PV power output and solar irradiance as a function of cell temperature changes is commonly used, derived from the efficiency relation to cell temperature from (19), and is given as follows [14], [15]:

$$P_{pv} = \frac{I_C}{I_{C,ref}}P_{pv,ref} \cdot \left[1 - \gamma_{ref}(T_{cell} - T_{ref})\right] \quad (26)$$

with $I_{C,ref}$ being the rated solar irradiance at STC.

Since (26) is in terms of cell temperature $T_{cell}$, it can now be used to replace $P_{pv}$ in (24) and obtain a function that can serve to calculate $T_{cell}$ at different values of incoming solar irradiance $I_C$ and ambient temperature $T_a$, which are ambient parameters whose profiles can be determined based on its array configuration and the theory of Section II. The result is a fourth-degree function for $T_{cell}$, expressed as follows:

$$0 = A\sigma\left[\left(\frac{1-\cos\Sigma}{2}\right)\varepsilon_a T_a^4 - \varepsilon_{cell}T_{cell}^4\right] + \alpha_{cell}I_C A - \frac{I_C}{I_{C,ref}}P_{pv,ref} \cdot \left[1 - \gamma_{ref}(T_{cell} - T_{ref})(T_{cell} - T_{ref})\right]. \quad (27)$$

Once the cell temperature is determined with (27) then the power output of the PV system can be obtained with (26).

It is important to note, however, that the linear relation of (26) is based on a regression model that in turn is based on testing made at different temperature points within a limited range of operating temperatures. For example, for a particular solar PV module suitable for space applications, the Spectrolab XTJ Prime, its data sheet specifies the temperature coefficient $\gamma_{ref}$ is relevant for a range between 15 °C and 125 °C [20]. For the scope of this paper, the linear relation of (26) is assumed to work for all the range of temperatures possible on the lunar surface (around -260 °C to 121 °C [6]). However, experimental procedures testing these PV modules under lunar-like conditions would need to be carried out in the future to determine a more precise relation between PV power output and cell temperature.

IV. RESULTS

Lunar ambient temperature $T_a$ is determined according to the horizontal irradiance profile assuming a direct relation with irradiance received by the lunar ground and its temperature. Additionally, PV collector irradiance $I_C$ is determined based on the array orientation configuration, which determines the effective angle of incidence $\theta$ the PV panels will experience at every time interval of the lunar year. Therefore, each array configuration will produce different power output profiles and hence, different net energy generation over the course of a lunar year.

The results in this section assume solar array systems composed of PV panels using the Spectrolab XTJ Prime module, which has the following parameters per module [6], [20]:

$P_{pv,ref} = 1.1368\ W$;  $I_{C,ref} = 1353\ W/m^2$;
$T_{ref} = 28\ °C\ (301.15\ K)$;  $\eta_{ref} = 30.7\%$;
$\gamma_{ref} = 0.1791\%1/K$;  $A = 0.0027\ m^2$;
$\alpha_{cell} = 0.8$;  $\varepsilon_a = 0.96$;
$\varepsilon_{cell} = 0.85$.

The overall PV system is assumed to be designed with a rated power of 1.1 MW. Therefore, an appropriate number of these XTJ Prime PV modules able to achieve this power rating is assumed.

Moreover, the temperature statistics vary depending on the latitude considered. In this case, for the south polar region near the Shackleton crater, assuming locations at the highest point of the crater's ridges, the minimum and maximum temperatures can then be obtained from [10]. For any other latitudes past 85°S, the temperature statistics found in [6] are used. TABLE I summarizes the temperature statistics used to determine ambient temperature profiles with (8) per latitude ranges.

TABLE I
Temperature statistics for different ranges of temperatures.

| Latitude Range [°S] | Max. Temperature [K] | Min. Temperature [K] |
|---|---|---|
| 0 ≤ L ≤ 25 | 394 | 94 |
| 25 < L ≤ 65 | 357 | 83 |
| 65 < L ≤ 85 | 224 | 41 |
| 85 < L ≤ 90 | 210 | 63 |

Additionally, Figs. 6 and 7 show horizontal irradiance and ambient temperature profiles, respectively, for three different regions: equatorial (0°), tropical (45°S), and polar (90°S). These ambient temperatures correspond to the lunar ground temperatures and are obtained by mapping horizontal irradiance using (8), and the statistics of TABLE I; hence they follow the same pattern.

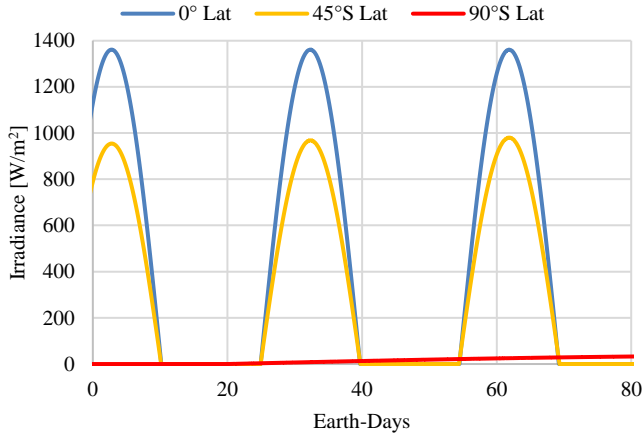

Fig. 6: Horizontal irradiance received by the lunar ground at different regions.

While the ambient temperature does have an influence on PV cell temperatures, the cell temperature will not be the same as the ambient temperature, and in most cases, it may not even be near the same temperature because some configurations seek to maximize solar irradiance exposure, which can heat up the PV panels to substantially higher temperatures than the ground, since it may be receiving irradiance at less direct angles, such as the case in the polar regions. This can be noticed in Figs. 8, 9, and 10, which show PV cell temperatures for the four different array configurations for equatorial, tropical, and polar regions respectively.

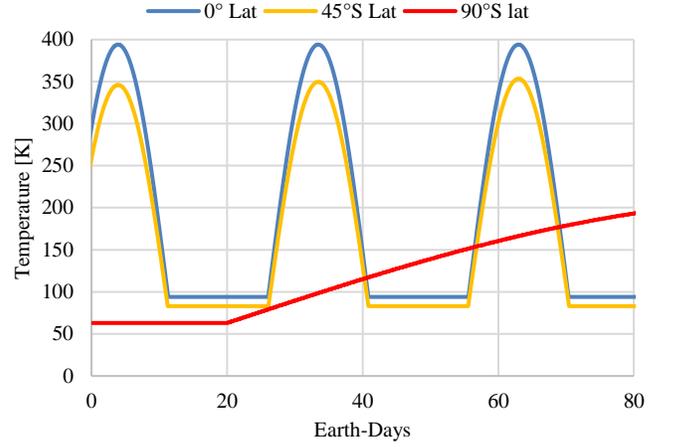

Fig. 7: Lunar ambient temperatures at different regions.

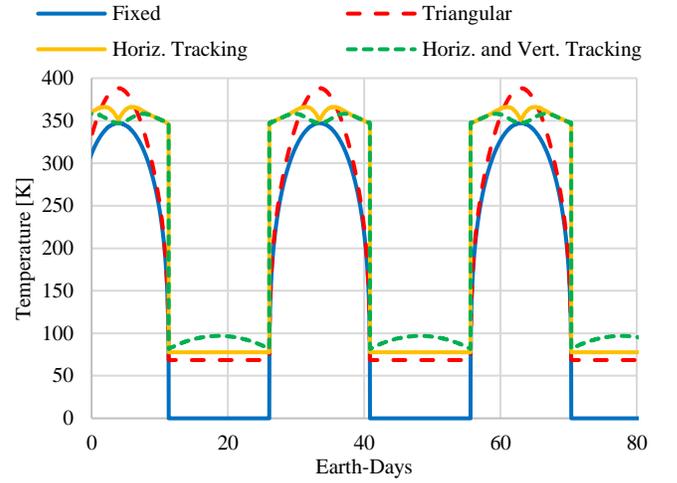

Fig. 8: PV cell temperatures at equatorial regions (0°).

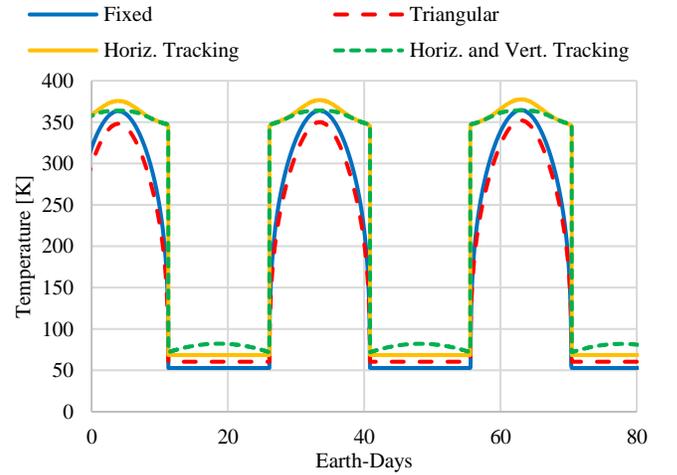

Fig. 9: PV cell temperatures at equatorial regions (45°S).

The configurations that incorporate tracking show how the PV cell temperature during daytime remains well over 300 K (27 °C) even at latitudes that exhibit lower ambient temperatures. This occurs as a result of their optimized angle of incidence that increases their collected irradiance as opposed to the lower horizontal irradiance the ground is receiving. As a result, the PV power output linear regression model represented by (26) can therefore be used with confidence for the XTJ Prime

PV modules, since their temperature is always within their operating range during daytime. For the fixed configurations, on the other hand, it can be used with confidence for the majority of the daytime as well, but at times near sunrise and sunset their cell temperature will drop below the stated operating limit (288 K).

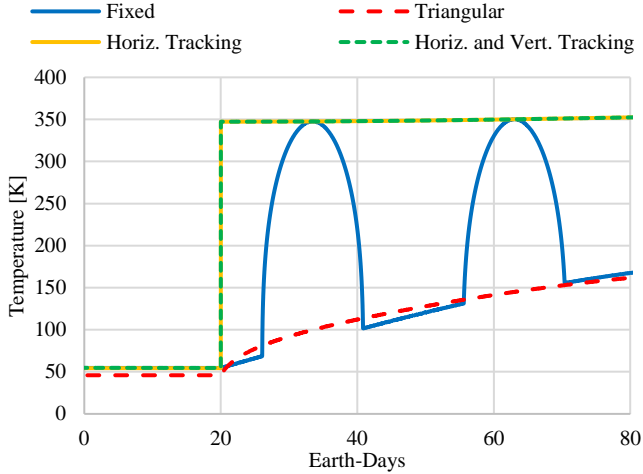

Fig. 10: PV cell temperatures at equatorial regions (90°S).

### A. Net Energy Generation Comparison

The results of the net solar energy generation in a lunar year (~347 Earth-days) are shown in Fig. 11 which displays a direct comparison between energy generation of different array configurations at different latitudes.

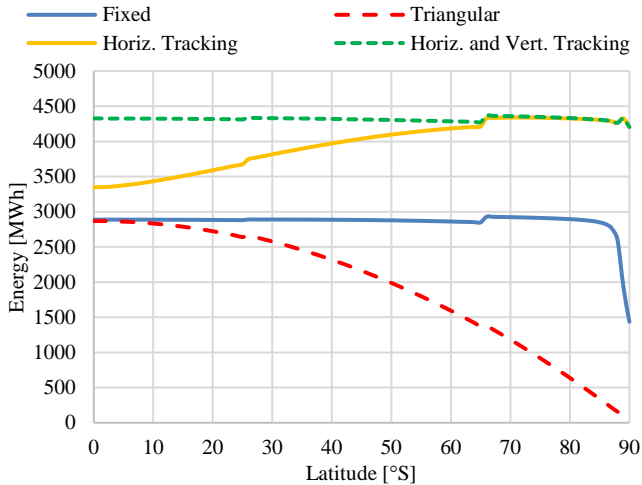

Fig. 11: Total solar energy generation per latitude for every array configuration.

The configuration that achieves the most optimal generation at all locations is obviously the array that integrates horizontal and vertical tracking of the solar azimuth and elevation angles. However, the amount of generation achieved by the array with horizontal tracking and a fixed tilt angle achieves close to the same amount as the array with horizontal and vertical tracking between 90°S and around 60°S. Therefore, it can be inferred that only enabling horizontal tracking and selecting an optimal fixed tilt angle based on the data of Fig. 5 is sufficient in near-polar regions, while on equatorial regions the benefit of the additional vertical tracking is more apparent.

For the two fixed configurations, the triangular array is able to achieve around the same generation as the fixed array only near the equator, with both also being relatively close to the energy generation of the horizontal tracking with fixed tilt array, but as soon as the location starts to move away from the equator, the benefit of implementing horizontal tracking is substantially larger. TABLE II shows a numeric summary of the capacity factor results for the equatorial, tropical, and polar regions on the moon. Based on these results, the horizontal tracking array may be recommended for polar regions since it is less complex than the horizontal and vertical tracking array and achieves essentially the same results. For tropical regions the horizontal and vertical tracking array provides a higher capacity factor of about 10.5%, however a cost-benefit as well as payload weight and volume analysis would be required to determine whether the more complex tracking system is worth the extra energy generation. For equatorial regions, on the other hand, the horizontal and vertical tracking array substantially surpasses all other configurations, and can be the recommended choice for these regions.

TABLE II
Solar PV system capacity factors by array configurations for three different lunar regions.

| Array Config. | Latitudes | | |
|---|---|---|---|
| | 0° | 45°S | 90°S |
| Fixed | 30.87% | 30.81% | 15.35% |
| Triangular | 30.66% | 23.09% | 1.12% |
| Horiz. Tracking | 35.78% | 43.16% | 44.93% |
| Horiz. and Vert. Tracking | 46.25% | 46.10% | 44.94% |

### B. Solar Power Output Profiles

The PV power output for each array configuration varies throughout lunar days, and also varies slightly as the lunar year goes by. For the same latitudes considered in TABLE II, Figs. 12, 13, and 14 display the power output profiles for each region respectively, for all four configurations during days that fall around the beginning of the lunar summer of its southern hemisphere.

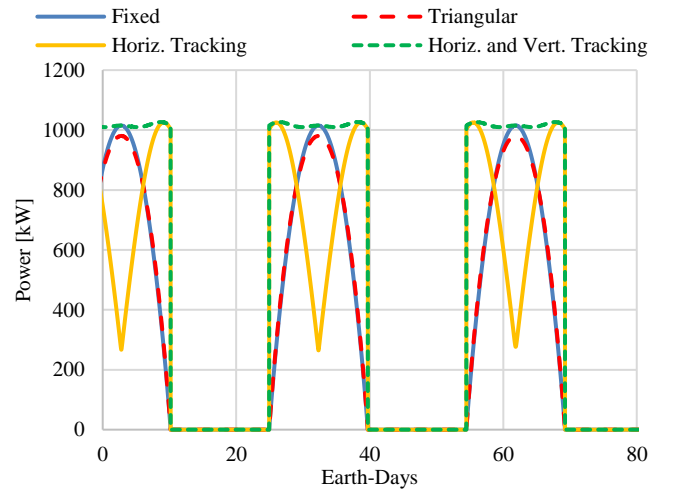

Fig. 12: PV power output profiles for equatorial regions (0°).

These results further evidence the benefit of solar tracking, with optimal horizontal and vertical tracking for equatorial regions and stand-alone horizontal tracking for polar regions. Moreover, while the triangular array could be said to achieve good performance for a less complex configuration near equatorial regions, it still performs very similar to the simple fixed array, however, its power output as well as total energy generation begins to drop as the location moves away from the equator, favoring those configurations with tracking enabled for these lunar surface applications.

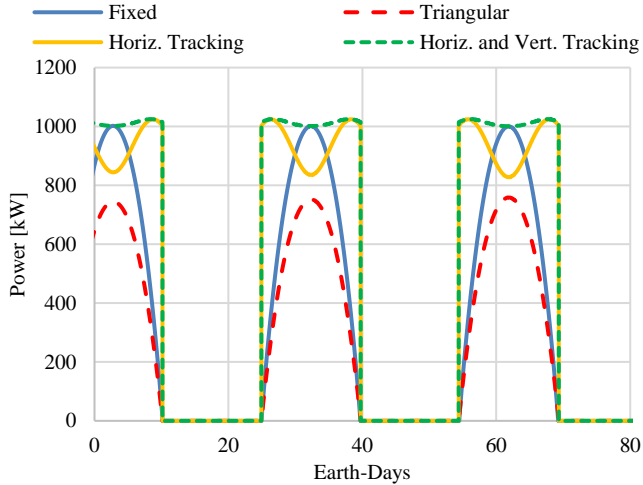

Fig. 13: PV power output profiles for tropical regions (45°S).

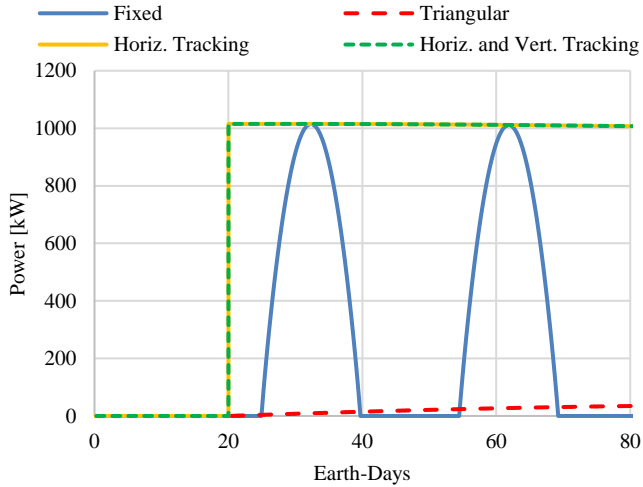

Fig. 14: PV power output profiles for polar regions (90°S).

## V. Conclusions

In consideration of the drastically different temperature ranges present on the lunar surface, as well as the very different ambient conditions compared to those on Earth, this paper proposes a PV power output mathematical model that can be used to determine power output profiles for a lunar year on the moon (~346 Earth-days) considering the efficiency and performance influenced by the PV cell temperatures. The model calculates power output as a function of PV cell temperature, which is in turn determined based on the lunar ambient temperatures as well as the net solar irradiance collected by the PV system. Additionally, this power output model is used to determine power output profiles for four different types of PV array orientation configurations at various lunar latitudes to compare their power outputs and net energy generations. These configurations are two fixed array types and two tracking-enabled array configurations. The fixed arrays include (i) a traditional fixed array with an inclination equal to the local latitude and (ii) a fixed triangular "tent-like" array configuration, while the tracking-enabled configurations include (i) an array with horizontal tracking and an optimal fixed tilt angle, and (ii) an array with both horizontal and vertical tracking.

The results indicate that even simply implementing horizontal tracking of the sun's location in the lunar sky provides substantial increases in solar power generation and the capacity factor. Moreover, between the two types of tracking-enabled arrays, the array with horizontal and vertical tracking outperforms the array with only horizontal tracking by a substantial difference in capacity factors near the equator, while near the south pole their performance difference is essentially negligible. Therefore, based on the results of this temperature-dependent PV power output model, a horizontal tracking array with an optimal fixed tilt can be recommended for polar regions given its performance and lower complexity, while for equatorial regions the inclusion of the additional vertical tracking can be justified by the substantial increase in generation.

ACKNOWLEDGMENT

This material is based upon work supported by the National Aeronautics and Space Administration (NASA) under award No. 80NSSC22M0300 through the Electric Power Research Institute (EPRI), with additional support from CenterPoint Energy (CNP). The authors would like to thank all the project team members from NASA's Johnson Space Center (JSC), EPRI and CNP for their valuable feedback.